**Title:** 2X8 Multilayer Microstrip Patch Array Antenna for C-Band Radar Micro-SAR.

**Name:** Pranoti S. Bansode

**Affiliation:** Assistant Professor

**Address:**
Department of Electronic Science,
Savitribai Phule Pune University,
Ganesh khind Road, PUNE-411007,
Maharashtra, INDIA.
**Mobile** - +91 8408875007
**Email:** Pranotisbansode@gmail.com; psb@electronics.unipune.ac.in

# 2X8 Multilayer Microstrip Patch Array Antenna for C-Band Radar Micro-SAR

Pranoti Bansode

Dept.of Electronics Science, SPPU.
pranotisbansode@gmail.com, psb@electronics.unipune.ac.in

**Abstract:** A Microstrip patch array antenna has been designed for 5.62 GHz on FR4 substrate having a dielectric constant of 4.4 and thickness is 1.6 mm. The antenna is used to design for use with a low cost imaging radar – MicroSAR. The antenna composed of an array having 8 X 2 patches could achieve the simulated gain of 17.5 dB with a bandwidth of 5%. The antenna consists of multilayer structure where two double side copper cladded substrates FR4 are used. Both the substrates are put in contact with each other and the common copper layer forms ground plane. The patch elements are placed on the top of one of the substrate and feed network is placed back side of the substrate. To increase the bandwidth of the antenna the height of the substrate is increased. Foam material is placed in between the two substrate to increase the height of the antenna. The corporate feed network is used to smoothly transferred the power from main port to the feed of antenna

## 1.0 Introduction

In order to acquire a high resolution large scale images of the earth, two dimensional images synthetic aperture radar is used. In all weather condition SAR provide unique images which represent the electrical and geometrical properties of the surface. For military and nonmilitary purposes such a wide rage radar has been used in applications like imaging, remote sensing and global positioning. SAR systems provide unique images representing the electrical and geometrical properties of a surface in nearly all weather conditions. Synthetic Aperture Radar is originated as an advanced form of side looking airborne radar (SLAR) which is the form of radar and it use of relative motion; between an antenna and its target region for obtaining finer spatial resolution [1]. SAR is implemented by mounting on a moving platform of spacecraft. The phased or patch array is placed on the radar which will send the RF beam towards the target and the beam will reflect toward the SAR. The reflected and the emitted wave will get compare and make the images of respective place. The image formation procedure produces an image that is two dimensional mapping.

Wasim Nawaz [1] designed a compact patch array antenna of 128 radiating elements. 4X4 sub array has been formed having total 16 elements and single fed constitute single sub array. Elements excited with corporate feed network where the inter element spacing of 0.6 λ. In this paper the dielectric material used is RT Duroid 5880 having dielectric constant 2.2 and thickness is 1.575mm. The loss tangent is 9X10^-4. The Advanced EM Simulator software is used to simulate the antenna. Simulated gain of the antenna is 17dB and return loss is around less than -50dB. Meandered transmission line at the center of the feeding network helps in better matching and hence improved bandwidth. For the 16 element the gain is small therefore total 128 radiating elements used to have gain of 26 dB. The total size of the antenna is 300 mm X 156mm. Meandered transmission line near the feeding point not only gives the phase difference of 180º between upper and lower sub arrays but also ensures better matching the side lobes are reduced. In this paper the size of the antenna is large as it contains 128 radiating elements and meandered transmission line actually decides the sub array matching and side lobe reduction.

Diptiman Biswas [2] in this paper author has designed Ku band array antenna with 4X4 rectangular microstrip elements for a distant point to point communication purpose. In this paper performance of single antenna, 1X2; 2X2; 2X4 and 4X4 array antennas have been computed and reported. IE3D electromagnetic simulation software is used for simulation of the antenna. A unique dc shorting technique has been proposed to enhance the antenna performance. The 16 element array antenna has been designed and constructed by using proximity coupled feeding technique. In this antenna the radiation from the microstrip patch array antenna is primarily because of the fringing field between the patch edge and ground plane. Microstrip patch antenna and the feed line shares a common ground plane. The RF energy from the feed line is coupled the radiating patch electromagnetically. Shorting Pin is connected to the patch and ground. The pin is moves from patch position and observe the readings. The resonant frequency changes as per the shorting via are moving. (Central location to outward open circuit) As the distance far from the patch frequency is shifts more towards higher range. VSWR is also changes as per the distance of shorting vias. The 4X4 Patch array antenna is fabricated having the bandwidth is 880MHz, gain of the antenna is 17.8 dBi. The HPBW is 17.8º in the elevation and 20º in the azimuth side. The frequency and VSWR is depends on the position of shorting via.

S. B. Chakrabarty [3] in this paper 3 element C band dual polarized linear array antenna has been designed and developed. The center frequency is of 5.3GHz. The single radiating element is used in the array configuration of stacked strip slot foam inverted patch (SSFIP). The design goal of the antenna is to achieve beam pointing by 10º, 3 dB beam width is 35º +/-2.5º. Cross polarization suppression better than -15dB and gain 11dBi. The spacing between the two consecutive elements should be less than λ. The ULTRALAM

2000 material is used as a dielectric material for both the patches. The substrates separating with supporting substrate 'Rohacell Foam' ($\varepsilon r=1.07$) and height is 4.8mm. The two feed networks below the ground plane which contains crossed slots have been structured according to corporate feed distribution. Ro4003 with $\varepsilon r=3.38$ has been chosen for two feed lines. The feed networks are separated by very thin substrate to enhance the port decoupling. Ground plane is placed at the rear side of the antenna to shield electromagnetically the associated electronic devices from the spurious back radiations. Dimensions of crossed slots have been chosen so that its resonant frequency is far beyond the operating bandwidth of the antenna. Thus minimizes the back radiations. The experimental results shows that the radiation pattern having side lobe levels below -15dB and -18 dB for both the planes. Elevation beam width is 35º at the center frequency. The measured gain is 10dB. The patches are made by using three different dielectric material which is quite difficult for the fabrication.

M. N. Jazi [4] designed and implemented the L Band aperture coupled patch antenna where aperture is designed to resonate near the resonant frequency of the patch. To improve radiation performance of the microstrip antenna four structures are proposed. The structure with air substrate has the maximum frequency bandwidth. The central frequency is 1060 MHz and the bandwidth is at least 250 MHz The thickness of patch substrate, h3 is chosen 20 mm due to the desired bandwidth. The feed substrate is RT/Duroid 5880 with thickness of h2 = 0.787 mm and relative dielectric constant of 2.2. FR4 is used as cover with thickness of h4 = 1 mm and $\varepsilon r4$ = 4.3. IE3D simulation software is used. The return loss of the antenna is less than −10 dB in the frequency range of 875 MHz–1150 MHz, i.e., the frequency bandwidth is greater than 25%. This aperture coupled microstrip antenna can be used as an element of microstrip array antennas in IFF (or SSR) systems. Here different dielectric materials have been used in four layers where the super substrate is used above the patch element for getting the maximum gain. Metallic plane is placed behind the antenna to reduce the back radiation power to -20dB. The metallic plane functions as a resonator and generates appropriate current distribution to eliminate the undesired radiation field. The distance between the feed line influences on the amplitude of the current distribution but the phase of the current distribution is controlled by the size of the back plane. The maximum gain and the radiation pattern is totally depends on the size of back plane.

Wonkyu Choi [5] presents the high gain patch array antenna using the superstrate layer where the layer affects its gain and resonant characteristics. Here two different types of antenna have been made, one is 2 X 8 antenna array with superstrate and 4X8 antenna array without superstrate. In single antenna structure it consist of two substrates, on the first substrate consist of one patch element and on the second substrate consist of second patch element. In between the two substrate there is foam material is present.

The lower patch and feeding line are on the substrate with εr = 2.17 and thickness = 0.508 mm, and the stacked patch on the thin film is supported by foam (εr = 1.06, 2 mm). Being very thin (0.04 mm) with respect to the wavelength, the film hardly affects radiation efficiency. The whole antenna structure is for getting high gain. The lower is 2mm being very thin compared to the wavelength radiation efficiency hardly affected. The resonance condition for a high gain is satisfied by adjusting the distance between superstrate and the radiating element. The distance D1 which is distance between patch element and the ground plane is electrically a quarter half wavelength. The distance between radiating element and the superstrate D2 is half wavelength. The gain of the antenna is improved by trial and error adjusting distance of D2.The thickness of the radiating element substrate is λg/2 which is very thick therefore problems such as high surface wave and cross polarization can occur.

Roopali Bhardwaj [6] paper presents the patch array antenna for the improvement in the gain. Defected ground structure (DGS) is used in the ground to improve the performance of the antenna. In this paper antenna consist of single substrate on FR4 material having dielectric constant 4.4. A 2X1 patch array antenna is designed. Frequency of operation is 7.642 GHz the S- parameter for this frequency is -21.993 dB. Gain of the antenna is 6.82 dB.

For such imaging radar the transmitting and receiving antenna plays very important role. For synthetic aperture radar the antenna having specifications like center frequency is at 5.8GHz.The antenna should have large bandwidth, it helps to take sufficient data for modification so it has bandwidth of 250MHz. It must have large gain around 18dB so resolution ability will be high. The beam width of the azimuth radiation pattern of the antenna should be narrow which around 8 ° is and elevation should be comparable wide around 25°. This paper presents the design of patch array antenna which is simulated, fabricated and tested for 5.8GHz. The antenna consists of 16 elements (8X2) on FR4 substrate having thickness 1.6mm. The patch dimensions are 18mm X 13mm. The substrate dimension is 244 mm X 107mm. Spacing between the two patches is 13 mm. The feed network is separated from patches itself for to avoid maximum spacing. In between the two substrate ground plane is present which is common for both patches and feed network.

## 2.0 Multilayer Patch Antenna Design

Microstrip patch antenna has the advantages of small size, various shapes, lower fabrication cost, support dual polarization, light in weight. Over this advantages microstrip Patch antennas also have limited functionality like spurious radiation exist in various patch antenna, offers low efficiency, offers low gain, higher level of cross polarization. To overcome this limited functionality, microstrip patch array antenna is better to use. In patch array antenna as the number of patch elements is decides the gain of antenna and the distance between the two patches decides the beam width of the antenna.

$$\text{Gain} = \frac{4\Pi Ae}{\lambda^2} \qquad \ldots\ldots [8]$$

For multilayer antenna design it consists of two substrates having same or different dielectric materials. as shown in the Fig. (1). Improve the bandwidth of the antenna the height of the substrate increases. The ground plane is placed between the two substrates. It will help to isolate the feed network from the radiating patches. Radiating patches and the feed networks are connecting by shorting vias. Foam as a dielectric material has been placed between the two substrate. These configurations have the advantage of increasing the antenna gain and also keeping out the losses (ohmic, conductor, line and dielectric) generated by the feed network in the overall antenna performance. Additionally, by installing the feed at the back of the patches; we are also conserving a big amount of space that is usually needed for the corporate feed network.

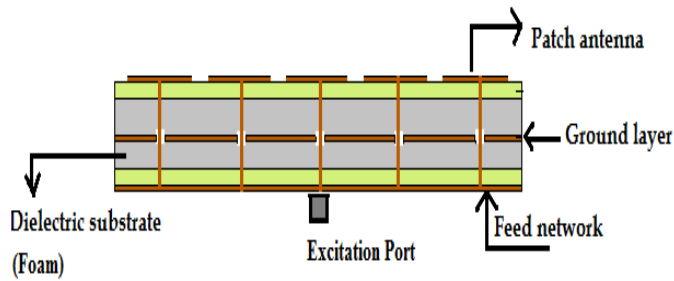

Fig.1. Side view profile of the antenna design

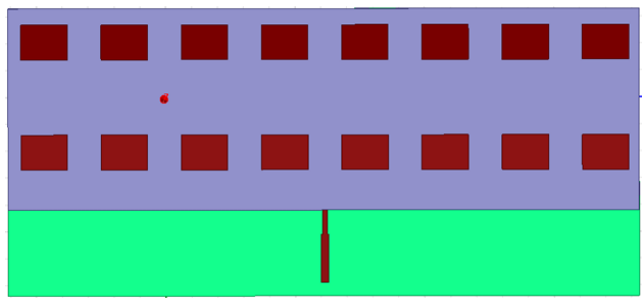

Fig.2. Simulated front view of the patch antenna.

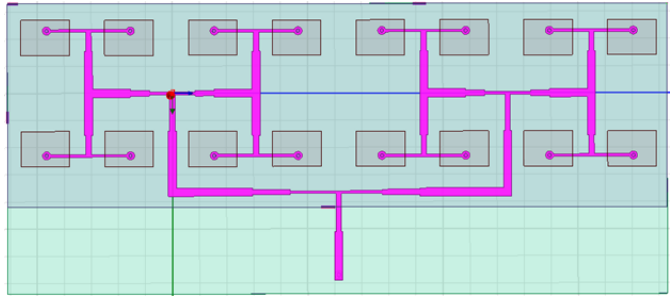

Fig.3. Simulated feed network view of the patch antenna.

**3.0 Feed Network Structure**

Corporate feed network is used to supply a uniformly distributed power to all sixteen patch elements. To ensure a smoother transition of impedance the transmission line should be tapered and no right angles for the feed bends. A quarter-wave transformer is also used wherever an impedance matching is needed. The Corporate feed network constructed is as displayed in Fig.3 the feed network is constructed from 100Ω impedance of the patches to achieve a shunt of 50Ω in between the two patches. Then, the 50Ω and 100 Ω feed lines have connected with 75 Ω feed lines to match the quarter wave impedance which is exactly after the 50 Ohm excitation line. Just before the main feed, the 50Ω feed line is tapered into the main feed that connects it with the SMA input power connector. First the feed network was simulated for 5.62 GHz and then same feed network was used with patch array antenna.

**4.0 Simulation Results**

The Microstrip patch array antenna is simulated in HFSS software. The simulated return loss of the antenna is observed in Fig.4. The figure shows a span of frequency from 5 GHz to 6 GHz and at the desired frequency of 5.63 GHz, a return loss value of -35 dB is achieved. From the same plot we could also see that the bandwidth of the antenna is around 112MHZ at -10 dB. Fig. 5 shows the VSWR which is 1.06.

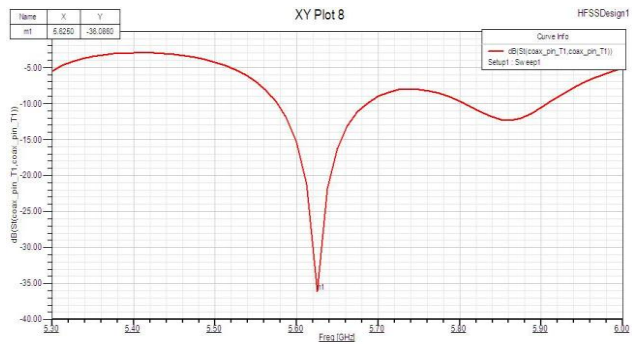

Fig.4. Simulated return loss of microstrip array antenna

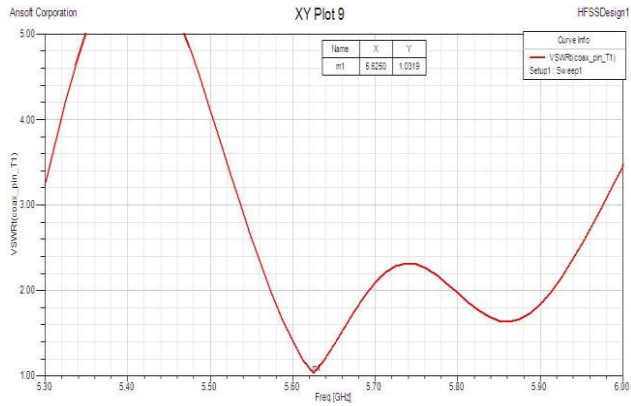

Fig.5. Simulated VSWR of microstrip array antenna of microstrip array antenna

The simulated two dimensional radiation pattern of the antenna could be observed in fig.6. The elevation gain of the antenna is around 8dB. The Fig 7. Shows the 3D polar plot of the radiation pattern.

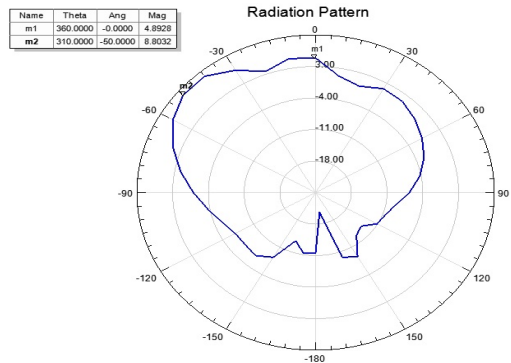

Fig.6. 2D Radiation Pattern of Microstrip patch array antenna

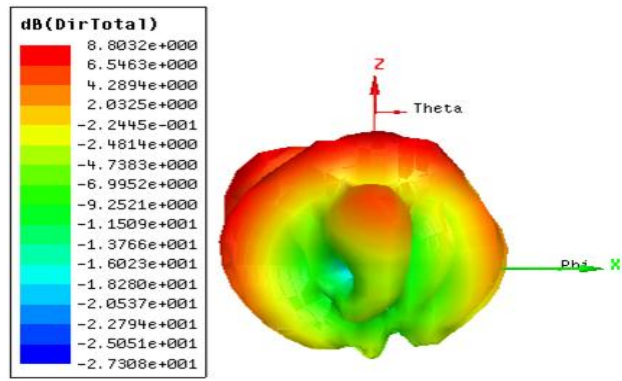

Fig.7. Gain of Microstrip patch array antenna

## 4.0 Experimental

The HFSS Software is used to design, model and simulate the antenna. The overall antenna dimension is 244 mm X 107 mm. The antenna consists of 16 elements (8X2) on FR4 substrate having thickness 1.6mm. The patch dimensions are 18mm X 13mm. Spacing between the two patches is 13 mm. Microstrip patch antenna with designed dimensions is fabricated on FR4 substrate using photolithography technique. The Foam material is placed between the two substrates for increasing the height of the antenna. The microstrip antenna radiates primarily because of fringing fields between patch edge and ground plane. Resonant frequency and return loss (lS11l) of the fabricated antenna are measured on Vector Network Analyzer (Agilent E5062A). The following fig.8 shows the fabricated 8 X 2 patch array antenna.

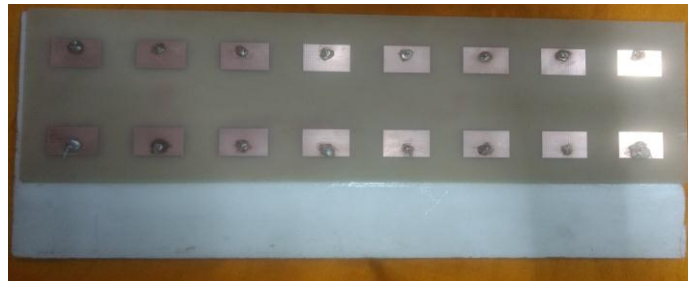

Fig.8. Top view of the Patch array antenna.

## 5.0 Fabricated Results

Microstrip patch antenna is fabricated by using photolithography technique. Antenna has been tested using vector network analyzer (VNA) N9923A.The experimental result of antenna exhibits the excellent impedance bandwidth of 79 MHz (from 5.788 GHz to 5.867 GHz). The measured return loss versus frequency of this array antenna has been shown in fig.9. The resonant frequency 5.8271GHz has return loss of -19.11 dB. A slight variation seems between the experimental and measured results. This may be due to the tolerance in manufacturing, uncertainty of the thickness and/or the dielectric constant of the substrate and lower quality of SMA connector (VSWR = 1.3), larger tangent loss = 0.0027 of the substrate and soldering effects of an SMA connector.

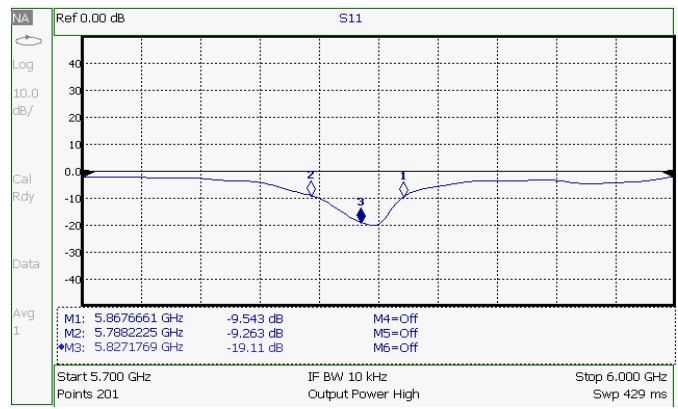

Fig.9. Experimental return loss of proposed antenna

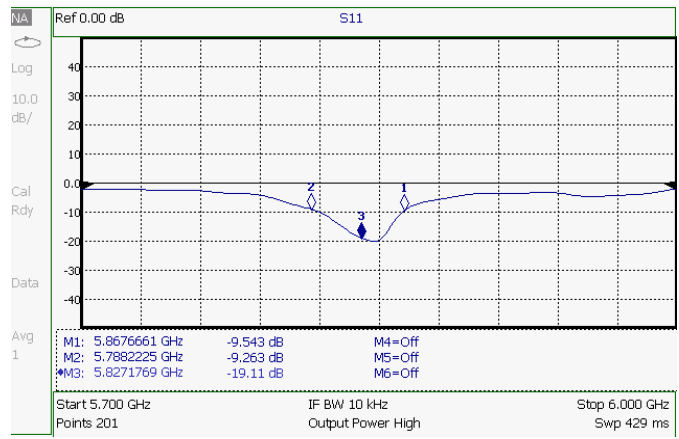

Fig.10. Experimental VSWR of proposed antenna

The current distribution pattern of the antenna is as shown in the fig 8. from current distribution pattern it seems that the current flowing in each part of the feed network and each element of the antenna is same and in phase.

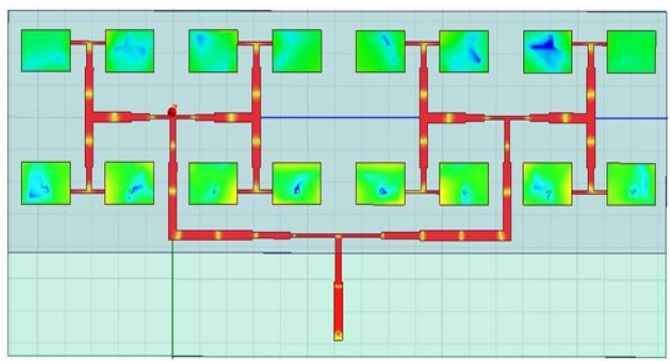

Fig.11. Simulated current distribution of patch elements and feed network

## 6.0 Conclusion

A 2X8 patch array antenna has been simulated, fabricated and tested. The simulated and fabricated results are not perfectly matched a slight variation seems between the experimental and measured results. The antenna is working for 5.8 GHz frequency. This may be due to the tolerance in manufacturing, uncertainty of the thickness and/or the dielectric constant of the substrate and lower quality of SMA connector (VSWR = 1.3), larger tangent loss = 0.0027 of the substrate and soldering effects of an SMA connector. The material used for the substrate is FR4 and Foam which is affordable and easily available in the market. The size of the antenna is not large so it is easy for the installation purpose also.

## 7.0 Acknowledgement

The author wish to thank Prof. S. AnanthaKrishnan, Adjunct Professor and Raja Raman Fellow, Department of Electronics Science, Savitribai Phule Pune University (SPPU) and Dr. S.A.Gangal, Emirates Professor, Department of Electronics Science, Savitribai Phule Pune University (SPPU) for providing fruitful technical discussions in preparing manuscript.